# Experimental demonstration of a novel bio-sensing platform via plasmonic band gap formation in gold nano-patch arrays

**Marco Grande[1,*], Maria Antonietta Vincenti[2], Tiziana Stomeo[3], Giuseppe Morea[1], Roberto Marani[1], Valeria Marrocco[1,5], Vincenzo Petruzzelli[1], Antonella D'Orazio[1], Roberto Cingolani[3], Massimo De Vittorio[3,4], Domenico de Ceglia[2], Michael Scalora[6]**

[1]*Dipartimento di Elettrotecnica ed Elettronica, Politecnico di Bari, Via Re David 200, 70125 Bari (Italy)*
[2] *AEgis Technologies Inc., 410 Jan Davis Dr., Huntsville 35806, AL – USA*
[3] *Italian Institute of Technology (IIT), Center for Bio-Molecular Nanotechnology, Via Barsanti, Arnesano (Lecce), 73010 Italy*
[4] *National Nanotechnology Laboratory (NNL), CNR-ISTITUTO DI NANOSCIENZE, Dip. Ingegneria dell'Innovazione, Università Del Salento, Via Arnesano, 73100 Lecce, Italy*
[5]*Istituto di Tecnologie Industriali ed Automazione ITIA-CNR, Via P. Lembo 38, 70125 Bari (Italy)*
[6] *Charles M. Bowden Research Center, RDECOM, Redstone Arsenal, Alabama 35898-5000 - USA*
[*]*grande@deemail.poliba.it*

We discuss the possibility of implementing a novel bio-sensing platform based on the observation of the shift of the leaky surface plasmon mode that occurs at the edge of the plasmonic band gap of metal gratings when an analyte is deposited on top of the metallic structure. We provide experimental proof of the sensing capabilities of a two-dimensional array of gold nano-patches by observing color variations in the diffracted field when the air overlayer is replaced with a small quantity of Isopropyl Alcohol (IPA). Effects of rounded corners and surface imperfections are also discussed. Finally, we also report proof of changes in color intensities as a function of the air/filling ratio of the structure and discuss their relation with the diffracted spectra.

## 1. Introduction

Recent progress in nanotechnology has lead to the fabrication of new optical sensors. Various technological approaches have been considered for the realization of devices: photonic crystals [1-5], ring resonators [6-7], evanescent field optical waveguides [8-9], metallo-dielectric structures [10-11] and microstructured fiber optics [12-13], just to mention a few. Optical sensing techniques based on surface plasmon coupling are indeed not new [14-16]. However, there is still room for improvement, especially when the detection of extremely small quantities of analyte is desired. Plasmonic sensing techniques reveal new features for the detection of molecular binding events or molecular structure changes. For example, this information is of primary importance in the study of biochemical reactions mediated by immuno-systems, tissues and whole cells [17]. Usually the target molecules are suspended in a generic biological liquid (analyte) that comes in contact with the active region of the optical sensor.

During the last few years different configurations of plasmonic sensors have been proposed, including the traditional Otto or Kretschmann arrangement that only deals with flat surfaces [18, 19], nanocavities [20, 21], nanoparticles [22-25], nanoclusters [26], nanopillars [27], nano-antennas [28], and/or nanorods [29] that aim to improve field localization in the sub-wavelength regions of the sensor. All these devices are designed to maximize the sensitivity, which in turn depends on the dynamics of the field in the specific geometry and on chemical interactions of the molecules within the active regions. In particular, nano-antennas [28, 30] manifest drawbacks because of the small size of the active region, which is usually restricted to the area between the two metallic arms. This distinctive feature reduces considerably the amount of target molecules that can be detected. Moreover, the presence of localized hot-spots in the gap between the metal rods limits the detection in the gap itself so that the sensitivity of the sensor becomes mainly a function of the distance between the arms of the antenna.

At the same time, the formation of bright and dark bands occurring at the onset of new diffraction orders of optical gratings dates back to the early 1900s [31, 32]. Following those experimental observations significant effort was devoted to trying to find a reasonable explanation of the phenomenon [33-35]: the modulation of the reflected spectrum occurs either at the appearance or fading of new spectral orders or when resonant-like anomalies are present in the grating. These two effects can occur either separately or simultaneously. In particular, the formation of such resonant modes corresponds to the excitation of leaky waves supported by the grating for specific spectral orders. The subject has enjoyed renewed interest after the observation of surface plasmon polaritons at metal/dielectric interfaces [18, 19].

The strategy to increase the sensitivity of plasmon-based sensors is thus to increase the area where strong field localization occurs. When dealing with metal gratings different resonant mechanisms can take place, each of them characterized by different field dynamics at the surface. Sub-wavelength slits can be periodically milled in a metal film giving rise to: (i) Fabry-Perot cavity modes whose localization resides mostly inside the slits (similar to nano-antenna hot-spots localization effects); (ii) leaky surface modes, whose localization involves the surface of the grating; (iii) hybrid modes, produced by the interference of Fabry-Perot cavity and surface waves [36]. Since field localization undergoes strong variations across the plasmonic band gap, caution should be exercised when exploiting such resonant mechanisms for sensing applications.

In this paper we report, for the first time to our knowledge, a theoretical discussion and an experimental demonstration of a novel bio-sensing platform that exploits the formation of a plasmonic band gap in two-dimensional arrays of gold nano-patches, together with the observation of a color change in the diffracted field when a small amount of IPA ($n$ = 1.37) is introduced on top of the grating.

## 2. Simulation Results

Figure 1 shows the sketch of the fabricated sample consisting of a two-dimensional array of gold patches periodically arranged on a Si substrate. The periodicities in the two directions are chosen so that $p_x = p_y = 630$ nm. Several samples with different aperture sizes were fabricated, while the thickness of the gold nano-patch is always equal to $w = 200$ nm. Simulations were performed by means of 2D and 3D Finite Element Method and a home-made 2D-FDTD, yielding nearly identical results.

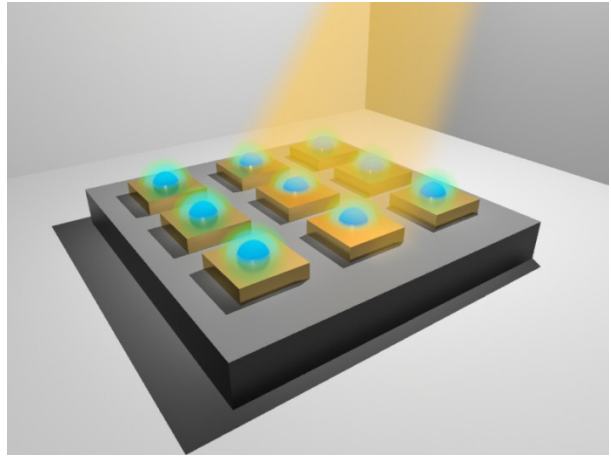

**Figure 1**. Sketch of the 2D gold arrangement on Silicon substrate (the cyan hemispheres represent the localized plasmonic resonances).

In previous papers [37, 38] an extensive comparison of the spectral response of 2D and 3D arrangements of gold patches was performed: periodic arrangements of square and rectangular patch arrays surrounded by air were studied under different polarization conditions, demonstrating how the structures similar spectral features whenever the periodicity in the two directions is the same. For this reason in the following we will compare the experimental results of the 2D arrangement with an equivalent 1D structure under TM illumination.

As outlined in more details elsewhere [36], metallic gratings where slits are periodically milled have several peculiar features in their spectra that depend mainly on the geometrical parameters of the structure. More specifically, three different states are distinguishable in the reflection spectrum across the plasmonic band gap: V, M and FP states, respectively. The V state corresponds to the excitation of a leaky mode that propagates on the metal grating and causes a sharp dip in the reflection spectrum; the M state corresponds to the coupling and back radiation of a surface plasmon and, as such, corresponds to a maximum in the reflectivity spectrum located at the wavelength $\lambda_{max}$ that satisfies the condition [39]:

$$q\lambda_{max} = p \left[\varepsilon_d \varepsilon_m(\lambda_{max})/(\varepsilon_d + \varepsilon_m(\lambda_{max}))\right]^{1/2} \qquad (1)$$

where $\varepsilon_d$ is the dielectric medium permittivity, $\varepsilon_m(\lambda_{max})$ is the real part of the dielectric function of the metal evaluated in correspondence of $\lambda_{max}$, and $q$ is an integer that defines the resonant order. Finally, the FP state is a hybrid state related to the availability of a Fabry-Pérot mode inside the cavity. This mode propagates in the metal-insulator-metal (MIM) waveguide that supports a transverse-electro-magnetic TEM-like guided mode only under specific circumstances.

Evidence of the above mentioned states is presented in Figure 2 (a), where the V state can be clearly identified on a two dimensional map where reflection is plotted as a function of both incident angle and wavelength. The simulated structure has the following parameters: *p = 630* nm, *w = 200* nm and the aperture of the slit *a = 120* nm. Here the wavelength range was limited to the visible range (400 – 700 nm) in order to compare the reflection values with the subsequent experimental results. The transmission map is obtained by varying the incident angle in the range 0° - 90°. The plot emphasizes how the resonance associated to the V state splits in two arms and linearly shifts with a variation of about 10 nm when the source is tilted by an angle as small as 1°. Furthermore, the absorption of gold between 400 nm and 500 nm suppresses all features within this wavelength range.

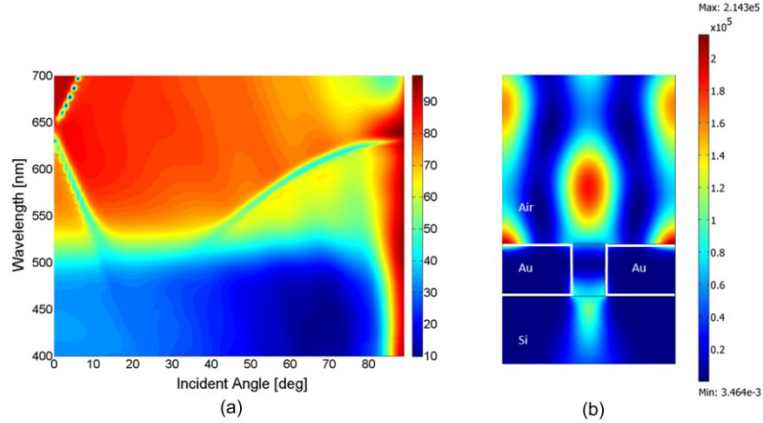

**Figure 2.** a) Reflection map when the incident angle is varied in the range 0° - 90°; b) Detail of magnetic field localization at 635 nm (corresponding to the V state); magnetic field intensity localizes on the whole grating surface effectively increasing the active area of the structure.

Figure 2 (a) also reveals that these geometrical parameters do not allow guided modes in the MIM waveguide for the wavelength range under investigation and, as a consequence, forbid the formation of hybrid modes related to the interference between cavity and surface modes. Moreover, we note that the full-width-half-maximum (FWHM) of the V state is about 5 nm wide at normal incidence, and it retains the same spectral width even for larger incident angles. We can thus speculate that this is probably the most favorable state for a sensing scheme. In fact as Figure 2 (b) shows, for normally incident TM-polarized light tuned to the V state the magnetic field embraces the entire grating, with a field enhancement up to 60 times larger than the incoming field [40].

We stress that the presence of the silicon substrate introduces resonances related to the silicon/metal interface. However, as one can easily infer from equation (1), the plasmonic band gap (centered at the M point) associated with the Si/metal interface is not visible in the spectra reported in Figure 2 (a) because it occurs at much longer wavelengths. Simultaneously, higher order plasmonic gaps associated with this interface are not visible in the range of interest because of the absorption of silicon in the frequency range below 500 nm.

## 3. Fabrication

The fabrication of the 2D arrangement of gold patches consists of three main steps: an electron beam lithography process followed by thermal evaporation and a lift-off process. The 2D pattern was written using a Raith150 e-beam lithography system operating at 30 kV. A bi-layer, composed of a 500 nm-thick Polymethyl methacrylate (PMMA)-MA and 200 nm-thick PMMA layers, was adopted as positive resist to ensure good resolution of e-beam writing (Figure 3 (a)). The key issue in the fabrication process is to achieve metal patches with well-controlled geometrical feature sizes. A preliminary dose-test was performed to define the optimum layout since the actual size of the pattern is influenced by the electron dose. A proximity error correction (PEC) was also applied to accomplish this target and the final dose was determined through Scanning Electron Microscope (SEM) inspections. Subsequently the sample was developed in a Methyl Isobutyl Ketone (MIBK) solution and then rinsed in an IPA-Methanol mixture. IPA solution was used as stopper for the development (Figure 3 (b)). Subsequently the 200 nm-thick gold layer was evaporated by means of a thermal evaporator with a current of 300 A, and a deposition rate of 2 Å/sec (Figure 3 (c)). In order to improve the adherence properties of gold on Silicon a 5 nm-thick chrome layer was deposited as adhesion layer. Finally, a lift-off process in an acetone bath was employed to remove the resist regions and reveal the final 2D array (Figure 3 (d)). The total thickness of the resist layer was chosen to facilitate the lift-off process thus avoiding the presence of resist in the slits.

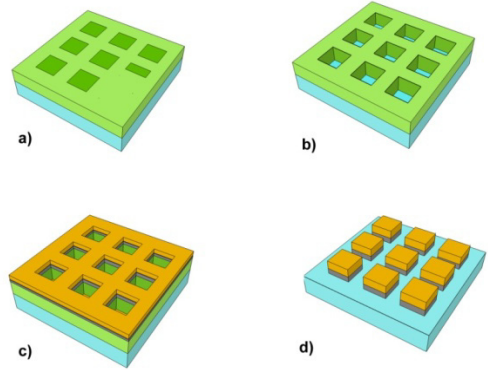

**Figure 3.** Sketch of the fabrication protocol of the 2D arrangement of gold patches on Silicon substrate (cyan): a) e-beam exposure of the bi-layer (green); b) MIBK and IPA-methanol mixture development; c) thermal evaporation of the chrome (grey) and the gold (yellow) layers, respectively and d) lift-off by means of an acetone bath.

Figure 4 shows the SEM image of the fabricated final device. As may be ascertained from the figure, the geometrical parameters obtained are almost identical to the nominal parameters of the structure simulated in Figure 1. The roughness of the evaporated gold layer was estimated by means of an Atomic Force Microscope (AFM) and a root mean square value of about 6 nm was obtained (Figure 4 (b)). This value is consistent with the experimental results since the roughness grows exponentially with the thickness of the metal film, as reported in [41]. Moreover, the structures have slightly rounded corners with a radius of about 20 nm. Several simulations were performed to assess the influence of fabrication defects on the spectral response of the structure. Rounded corners are indeed unimportant when monitoring the spectral position of the plasmonic band gap, since their features are mostly dependent on the periodicity of the array. The only effect of rounded corners having radii greater than 50 nm is a slight shift in the V state position, which is compatible with an effective change of the periodicity experienced by the incident light. On the other hand, the main effect of roughness is broadening of the V state bandwidth, as we will show in what follows. In order to compare this structure with other proposed plasmonic-based sensors, we measured the active area involved in the sensing process: Figure 4 (a) highlights that period and slit size are equal to 630 nm and 121 nm, respectively, which means that the active area of the structure is ~ 65 %, significantly higher than the active area usually exploited for plasmonic sensing via nanoparticles and nano-antennas [22-30].

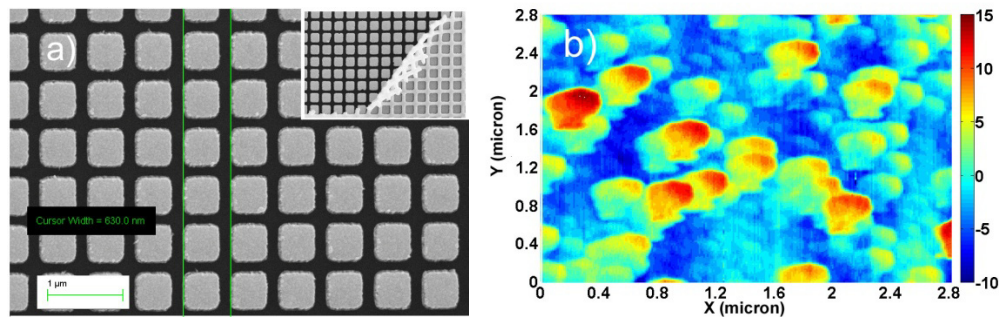

**Figure 4.** (a) SEM image of the fabricated device (inset: lift-off process); (b) roughness on a total area of 2.8 x 2.8 μm² estimated by means of Atomic Force Microscope.

## 4. Optical Characterization

The device was characterized by means of the optical setup sketched in Figure 5. A white light lamp, filtered in the 400 nm – 700 nm range, was focused on the sample by means of a low numerical-aperture infinity-corrected microscope objective (5x, $NA = 0.10$). The reflected light is collected by an aspherical fiber lens collimator and sent to an optical spectrometer (HR4000 from Ocean Optics) through a multimode optical fiber. The HR4000 has a 3648-element CCD-array detector that enables optical resolution as precise as about 1 nm (Full Width Half Maximum, *FWHM*) in the range of interest and provides an integration time up to 65 seconds.

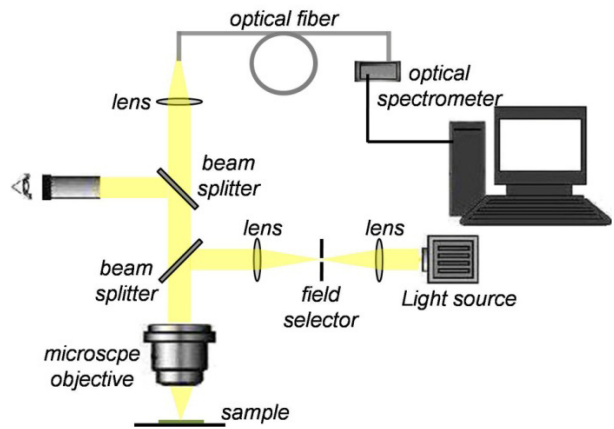

**Figure 5.** Sketch of the optical setup.

The experimental result at normal incidence is compared with the simulated spectrum in Figure 6. The reflected light on the bare Si surface was used as the reference signal for the normalization of the reflection spectra of the metallic patches. The measured spectrum has a dip located around 634 nm (blue line in Figure 6), while the expected resonance obtained by the numerical analysis is equal to 635 nm (red line in Figure 6). This extremely small shift and the broadening of the measured spectra are compatible with the measured roughness and rounded corners of the sample. In fact the broadening effect was verified by means of a FEM simulation where a random roughness having an average value of 6 nm was introduced on the surface of the grating and on the walls of the slits between the gold nano-patches (green line in Figure 6).

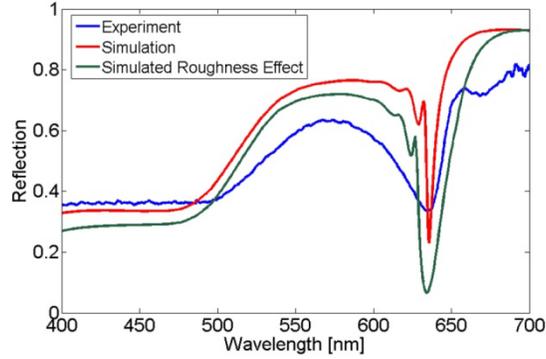

**Figure 6.** Measured (blue) and simulated (red) reflection spectra. Green solid line shows the effects of an average roughness along the grating and inside the metal walls similar to the measured roughness shown in Figure 4 (a).

The enlargement of the measured spectrum may come about also as a result of the Airy pattern due to diffraction or scattering of the light as it passes through the circular back aperture of the objective. For monochromatic light this phenomenon leads to the formation of concentric light and dark circles where the central one is usually defined by an Airy disk and contains about 84 % of the total power. In our case the same considerations can be applied to the superposition of the different wavelengths composing the white light. Moreover, the resonance of the metallic 2D patches is highly sensitive to the variation of the incident angle of the source (see Figure 2 (a)). Therefore the overall resonance can also be interpreted as a weighted superposition of the resonances in the angle range *–sin (NA)*, *+ sin (NA)*. Since the NA of the microscope objective is equal to 0.10, the device is illuminated by light in the cone -6°, + 6°, hence the resonance experiences broadening leading to a *FWHM* of about 20 nm.

As mentioned above, the V state is very favourable to perform sensing measurements. Since this state involves the whole surface of the grating even an extremely small change in the refractive index of the material covering the sample will cause a detectable shift in the V state. As an example of the potential sensing performance of the structure, we simulated and compared the reflected spectra at normal incidence for the structure depicted in Figure 1 when air (blue line) and a material with $n$ = 1.001 (red line) cover the gold nano-patch array grown on a Silicon substrate, respectively. Figure 7 shows a shift of 1 nm in the spectra. This shift corresponds to a sensitivity *S* equal to 1000 nm/RIU with a theoretical *FWHM* equal to 4.5 nm. The resulting Figure of Merit (*FOM*) *S/FWHM* is equal to 222 RIU$^{-1}$.

Another method to reveal the presence of a chemical compound on the top of this plasmonic structure is to look at the light diffraction in far field. A microscope in dark field configuration (DF) was used to investigate the scattering behaviour of the periodic structure of Figure 1. Under these conditions, the device is illuminated by light that impinges on the sample at an oblique angle equal to about 14° (10x, *NA* = 0.25). Figure 8 (a) shows the superposition of the experimental scattering spectrum (solid black line) of the sample when illuminated by the white light lamp and the diffraction spectrum related to the order *m* = -1 of the 2D arrangement (dashed black line). The comparison proves that the performance of the device resembles the behaviour of the ideal structure with slight differences that could be due to the presence of roughness and rounded corners in the real pattern. Moreover, the plot shows that scattering is mainly located in the range 400 nm – 520 nm while it is completely negligible for wavelengths longer than 520 nm where diffraction can be only associated with the zeroth diffraction order. This diffraction order is completely suppressed by the microscope.

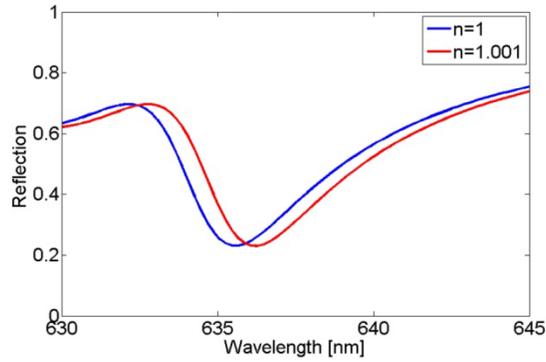

**Figure 7.** Simulated spectral shift of the V state when air (blue line) or a material with *n* = 1.001 (red line) cover the gold nano-patches. A shift of 1 nm is detected for such a small change of the refractive index.

Figure 8 (b) shows a picture of the same sample captured by a digital camera: it is clearly evident that only the periodic pattern contributes to the scattering process while the bare Silicon surface appears completely black due to its non-scattering nature and flat surface. The color of the 2D metallic patches appears uniform and homogenous over the entire structure and corresponds to the color of the maximum value. The conversion of the peak at 461 nm color line corresponds to the RGB color scale counterpart. The comparison confirms that the captured color and the corresponding conversion identical. In the same picture, a white line may be recognized due to write field dimension related to the electron beam lithography step. At the same time, small particles deposited on the sample scatter light and introduce small spots in the black background. The background scattering generated by these particles is negligible with respect to the periodic structure signal as confirmed by the scattering measurement.

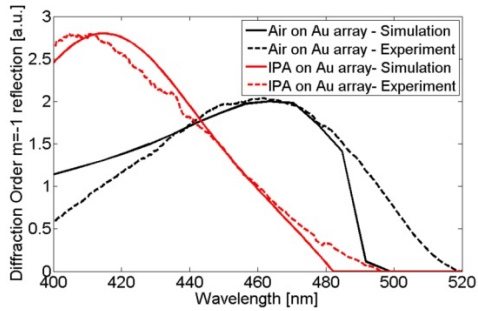
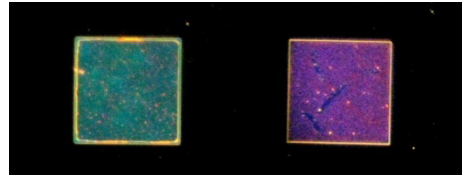

(a)　　　　　　　　　　　　　(b)　　　　(c)

**Figure 8.** a) Measured and simulated scattering spectra when the device is covered by air (black) and IPA (red), respectively. Picture captured by a digital camera when the device (with a dimension of 100 μm x 100 μm) is covered by b) air and c) IPA alcohol, respectively.

The same measurement was repeated after pouring the IPA solution on the sample: scattering from the structure experiences a blue shift consequent to the change of the refractive index of the top interface. All the considerations made for the air case are still valid also for the IPA solution. Figure 8 (a) reports measured and calculated diffraction spectra, respectively solid and dashed red lines, for the structure with IPA solution on top: in this case the theoretical and experimental results almost completely overlap. Figure 8 (c) shows the corresponding picture captured by the digital camera that reveals modification of the color from cyan to violet. We note that since the IPA is a very volatile alcohol the evaporation process can also be monitored through the change of the color of the sample, where cyan and violet regions alternate as the observation time is increased. This is due to the small thickness of the IPA solution on the sample that becomes IPA-free after a few minutes. This time interval is sufficient to measure the reflection spectra.

Several samples with different air filling ratios were also fabricated. The percentage of air in the metal structure obviously alters the scattering even when the period is fixed at $p = 630$ nm. In particular, we change the aperture $a$ in the range of 60 nm up to 120 nm in steps of 15 nm. Figure 9 shows the dark field measurement of five different arrays with different air filling ratio (9.5 %, 11.9 %, 14.3 %, 16.7 % and 19 %). The picture clearly demonstrates that the color shifts from dark cyan to light cyan, from left to right, respectively. It is worth noting that the same behavior is observed regardless of the position of the arrays in the optical field of the microscope. Simulated reflected diffraction orders are reported on the bottom of Figure 9. The simulated results exhibit a decrease in the cyan component of the spectrum with the decrease of the aperture size, which is consistent with what one can observe in the DF microscope.

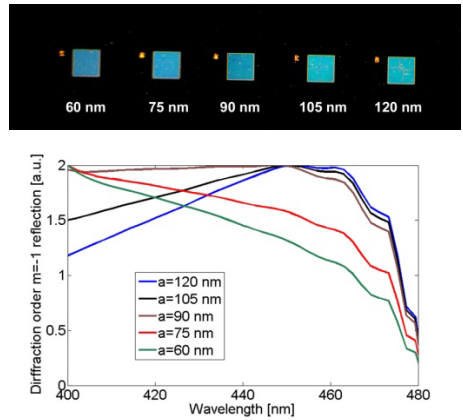

**Figure 9.** Top: scattering behavior when the aperture $a$ is varied from 60 nm to 120 nm. Bottom: Simulated reflection for $m = -1$ diffraction order.

## 5. Conclusion

In conclusion, we have reported on the design, fabrication and characterization of 2D gold metallic patches on silicon. In particular, the numerical analysis reveals that these periodic structures exhibit very narrow resonances corresponding to the formation of leaky modes associated with surface plasmons of the air/metal interfaces (state V). This condition is extremely sensitive to incident angle and refractive index variations at the same interface. Because of its finesse and the field localization profile, the V resonance may be recognized as the best tuning conditions for bio-sensing applications. The experimental measurements confirm the spectral behavior predicted by the simulations and emphasize that roughness does not noticeably affect the performance of the device. Moreover, we have experimentally demonstrated how the brightness of the device can be altered by varying the aperture size and keeping the periodicity constant.

The potential sensitivity of the device may be derived by evaluating the spectral shift when an extremely small change of the refractive index is induced at the top surface of the sensor, leading to sensitivity values up to 1000 nm/RIU and a corresponding Figure of Merit (*FOM*) of 222 RIU$^{-1}$ (*FWHM* of the resonance is only 4.5 nm) that corresponds to a significant advance in the improvement of plasmonic sensor devices. Moreover, we have shown a new, more intuitive path to detect variations of refractive index from color changes. We compared the diffracted spectra of the same sample with air and an IPA solution on top, registering a significant change in the sample color. This observation can be also quantified in terms of sensitivity, leading to $S$ = 121 nm/RIU and $FOM$ = 6 RIU$^{-1}$ (*FWHM* of the measured spectra is equal to 20 nm). We stress that, while the *FOM* obtained by observing the diffracted spectra is significantly smaller than that achieved by monitoring the shift of the V state, the color observation method is an easy, instantaneous way to ascertain the presence of chemicals or other foreign substances on the surface of a device without resorting to any additional instrumentation. Moreover, such observation method allows establishing numerous other properties of the foreign compound such as its thickness or evaporation time, paving the way for new, yet unexplored plasmonic sensor devices.